\documentclass[10pt,a4paper,twocolumn,superscriptaddress,aps,prd,longbibliography,nofootinbib]{revtex4-2}
\usepackage[utf8]{inputenc} 
\usepackage[T1]{fontenc}    
\usepackage{lmodern}        

\usepackage[modulo]{lineno}
\usepackage{amsmath}
\usepackage{amsfonts}
\usepackage{amssymb}
\usepackage{bbold}
\usepackage{graphicx}
\usepackage[usenames,dvipsnames]{color}
\usepackage{float}
\usepackage{multirow}
\usepackage{soul}
\usepackage{hyperref}
\usepackage{ulem}

\usepackage[nohyperlinks, printonlyused, nolist]{acronym}

\usepackage[range-units=single,separate-uncertainty]{siunitx}
\usepackage{verbatim} 

\graphicspath{ {./images/} }

\usepackage{blindtext}

\begin{document}

\author{Lukas Gehrig}
\thanks{These authors have contributed equally}
\affiliation{Physikalisches Institut, Universit\"at W\"urzburg, D-97074 W\"urzburg, Germany}
\affiliation{W\"urzburg-Dresden Cluster of Excellence ctd.qmat, Universit\"at W\"urzburg, D-97074 W\"urzburg, Germany}

\author{Cedric Schmitt}
\thanks{These authors have contributed equally}
\affiliation{Physikalisches Institut, Universit\"at W\"urzburg, D-97074 W\"urzburg, Germany}
\affiliation{W\"urzburg-Dresden Cluster of Excellence ctd.qmat, Universit\"at W\"urzburg, D-97074 W\"urzburg, Germany} 

\author{Erica Fragomeni}
\thanks{These authors have contributed equally}
\affiliation{Department of Physics, Sapienza University of Rome, Piazzale Aldo Moro 5, 00185, Rome, Italy}

\author{Simone Sotgiu}
\thanks{These authors have contributed equally}
\email{e-mail: simone.sotgiu@physik.rwth-aachen.de}
\affiliation{JARA-FIT and 2nd Institute of Physics, RWTH
Aachen University, 52074 Aachen, Germany}
\affiliation{Peter Gru\"unberg
Institute (PGI-9), Forschungszentrum J\"ulich, 52425 J\"ulich, Germany}

\author{Stefan Enzner}
\affiliation{W\"urzburg-Dresden Cluster of Excellence ctd.qmat, Universit\"at W\"urzburg, D-97074 W\"urzburg, Germany}
\affiliation{Institut f\"ur Theoretische Physik und Astrophysik, Universit\"at W\"urzburg, D-97074 W\"urzburg, Germany}

\author{Tommaso Venanzi}
\affiliation{TUD Dresden University of Technology, Dresden, Germany. }

\author{Bing Liu}
\affiliation{Physikalisches Institut, Universit\"at W\"urzburg, D-97074 W\"urzburg, Germany}
\affiliation{W\"urzburg-Dresden Cluster of Excellence ctd.qmat, Universit\"at W\"urzburg, D-97074 W\"urzburg, Germany}

\author{Kilian Strau\ss}
\affiliation{Physikalisches Institut, Universit\"at W\"urzburg, D-97074 W\"urzburg, Germany}
\affiliation{W\"urzburg-Dresden Cluster of Excellence ctd.qmat, Universit\"at W\"urzburg, D-97074 W\"urzburg, Germany}

\author{Jonas Erhardt}
\affiliation{Physikalisches Institut, Universit\"at W\"urzburg, D-97074 W\"urzburg, Germany}
\affiliation{W\"urzburg-Dresden Cluster of Excellence ctd.qmat, Universit\"at W\"urzburg, D-97074 W\"urzburg, Germany}

\author{Martin Kamp}
\affiliation{Physikalisches Institut, Universit\"at W\"urzburg, D-97074 W\"urzburg, Germany}
\affiliation{Physikalisches Institut and R\"ontgen Center for Complex Material Systems, D-97074 W\"urzburg, Germany}

\author{Elena Stellino}
\affiliation{Department of Basic and Applied Sciences for Engineering, Sapienza University of Rome, 00185 Rome, Italy}

\author{Paolo Postorino}
\affiliation{Department of Physics, Sapienza University of Rome, Piazzale Aldo Moro 5, 00185, Rome, Italy}

\author{J\"org Sch\"afer}
\affiliation{Physikalisches Institut, Universit\"at W\"urzburg, D-97074 W\"urzburg, Germany}
\affiliation{W\"urzburg-Dresden Cluster of Excellence ctd.qmat, Universit\"at W\"urzburg, D-97074 W\"urzburg, Germany}

\author{Simon Moser}
\affiliation{Physikalisches Institut, Universit\"at W\"urzburg, D-97074 W\"urzburg, Germany}
\affiliation{W\"urzburg-Dresden Cluster of Excellence ctd.qmat, Universit\"at W\"urzburg, D-97074 W\"urzburg, Germany}
\affiliation{Experimentalphysik IV - AG Oberflächen, Ruhr-Universität Bochum, 44801 Bochum, Germany}

\author{Christoph Stampfer}
\affiliation{JARA-FIT and 2nd Institute of Physics, RWTH
Aachen University, 52074 Aachen, Germany}
\affiliation{Peter Gru\"unberg
Institute (PGI-9), Forschungszentrum J\"ulich, 52425 J\"ulich, Germany}

\author{Giorgio Sangiovanni}
\affiliation{W\"urzburg-Dresden Cluster of Excellence ctd.qmat, Universit\"at W\"urzburg, D-97074 W\"urzburg, Germany}
\affiliation{Institut f\"ur Theoretische Physik und Astrophysik, Universit\"at W\"urzburg, D-97074 W\"urzburg, Germany}

\author{Ralph Claessen}
\affiliation{Physikalisches Institut, Universit\"at W\"urzburg, D-97074 W\"urzburg, Germany}
\affiliation{W\"urzburg-Dresden Cluster of Excellence ctd.qmat, Universit\"at W\"urzburg, D-97074 W\"urzburg, Germany}

\author{Leonetta Baldassarre}
\email{e-mail: leonetta.baldassarre@uniroma1.it}
\affiliation{Department of Physics, Sapienza University of Rome, Piazzale Aldo Moro 5, 00185, Rome, Italy}

\date{\today}


\title{Resonantly-enhanced Raman response in graphene-capped bismuthene on SiC}
\maketitle

\noindent \textbf{Two-dimensional quantum spin Hall insulators based on atomic monolayers offer a promising route toward dissipationless electronics, yet their practical use is often limited by environmental instability. Encapsulating the system with a graphene capping layer has been shown to be a reliable method to prevent oxidation and degradation. However, the confirmation of a successful encapsulation still relies on ultra-high vacuum techniques, that considerably slow the process.
Here, we present an \textit{ex situ}, rapid, non-destructive and spatially resolved Raman characterization of graphene-capped bismuthene, a honeycomb monolayer of Bi on SiC. A pronounced Raman scattering peak at $\sim$122~cm$^{-1}$ is identified as the $E_{2g}$ phonon of bismuthene, via a comparison with density functional perturbation theory calculations. We use excitation-energy and polarization-dependent Raman measurements to enable an unambiguous assignment of the spectral features. Tuning the excitation energy close to the excitonic transition in pristine bismuthene, we observe a strong enhancement of the Raman response and the emergence of additional scattering peaks. In this regime, higher-order phonon features, as well as interfacial modes between bismuthene and the SiC substrate, become visible, suggesting the involvement of resonant scattering processes. Our results establish Raman micro-spectroscopy as a versatile tool for probing graphene-protected quantum materials, providing access to lattice dynamics and interlayer coupling.}

\section*{}

Two-dimensional (2D) quantum spin Hall insulators (QSHIs) represent a central platform for exploring topological phases of matter and their potential for dissipationless electronics \cite{Weber2024,Trauzettel2013,Kou2017}. In these systems, a bulk band gap coexists with topologically protected edge states that enable spin-polarized transport immune to backscattering \cite{Bernevig2006,Kane2010}. Among the various material realizations, bismuthene, a honeycomb monolayer of bismuth epitaxially grown on SiC(0001), stands out due to its exceptionally large ($\ge800\,$meV) nontrivial band gap \cite{Reis2017}, which is the largest reported for any QSHI and thus enables operation at elevated temperatures.

\begin{figure*}[t!]
\includegraphics[width=\linewidth, keepaspectratio]{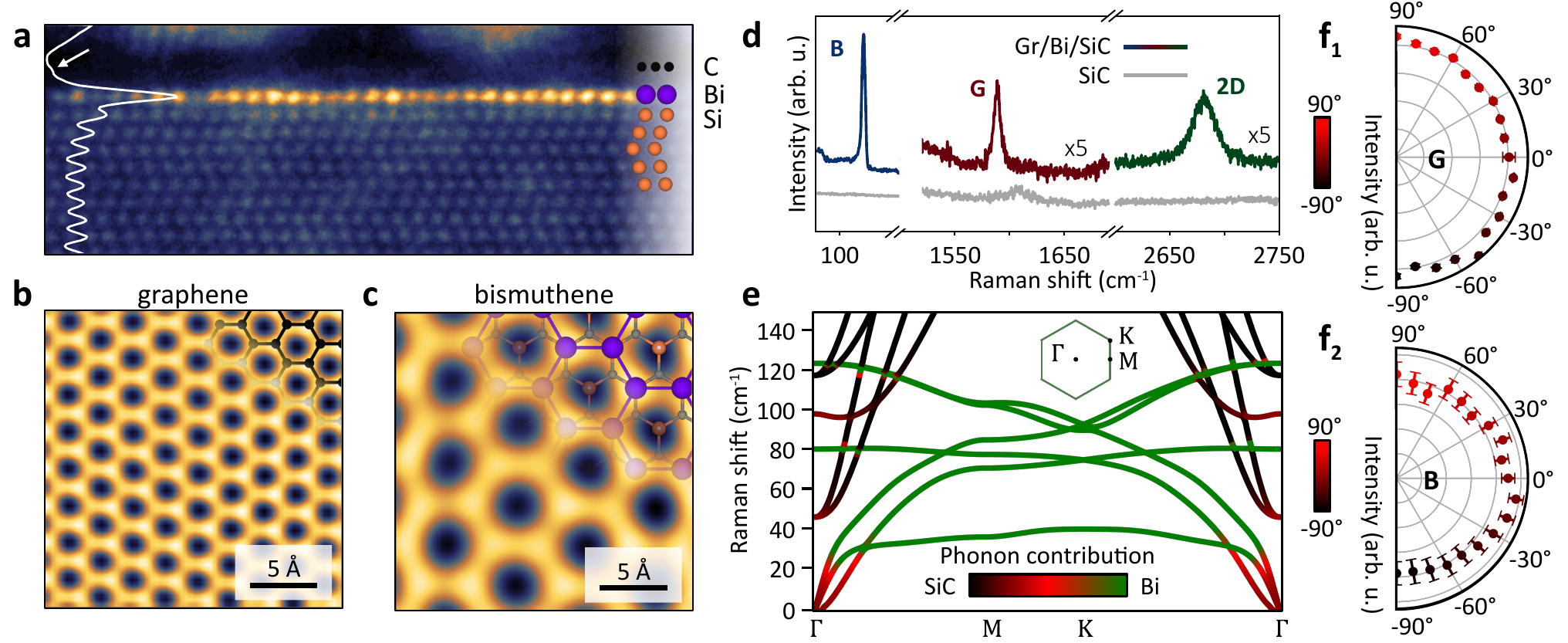}
\caption{\textbf{Raman signature of intercalated bismuthene.} \textbf{a} STEM image revealing a single monolayer of intercalated bismuth atoms, with the adsorption site located at the T1 position. \textbf{b} and \textbf{c} show STM images of graphene and bismuthene, respectively (see text for details). Both structures exhibit a honeycomb lattice, but with different lattice constants. \textbf{d} Raman spectrum showing the characteristic B peak of bismuthene (blue), as well as the G and 2D peaks of graphene (red and green, respectively). For comparison, a reference spectrum of the bare SiC substrate is shown in grey. \textbf{e} DFPT-calculated phonon dispersion of bismuthene on SiC, revealing a purely bismuth derived mode at $\Gamma$ corresponding to the 122~cm$^{-1}$ peak. \textbf{f} Polarization dependence of the f\textsubscript{1} graphene G and f\textsubscript{2} bismuthene B peak.}
\label{fig:fig1}
\end{figure*}

Despite its remarkable electronic properties, bismuthene suffers from a critical limitation: its lack of stability under ambient conditions \cite{Gehrig2025}. This instability has so far hindered its integration into \textit{ex situ} experiments and device fabrication workflows, which are essential steps toward technological applications. A promising route to overcome this challenge is the encapsulation of bismuthene on SiC beneath graphene \cite{Gehrig2025}. Similar to other intercalated 2D system, the graphene capping layer acts as an effective, atomically thin barrier that protects the underlying quantum material from environmental degradation, while preserving its structural and electronic properties  \cite{Schmitt2024NatComm, Schmitt2024Raman,Ag,Pb1,Pb2,Au_Intercalation}. This approach opens the door to \textit{ex situ} characterization and scalable device processing \cite{Schmitt2024Raman}.

In this context, establishing rapid, non-destructive, and spatially resolved characterization techniques is crucial. Raman spectroscopy is uniquely suited for this task: it provides direct access to lattice vibrations, is experimentally fast and cost-effective, and can be performed under ambient conditions with sub-micrometer spatial resolution. While Raman spectroscopy has been extensively used to probe the quality, strain, doping, and defect density of graphene and other 2D materials \cite{venanzi2023probing,graf2007spatially,graziotto2024infrared,saito2016raman,Ga1,Ga2,zhang2015phonon,Wetherington_2021}, its application to encapsulated topological materials architectures remains largely unexplored \cite{Schmitt2024Raman}.

Here, characteristic Raman-active phonon modes of the bismuthene layer are identified, including a prominent in-plane optical mode analogous to the well-known Raman G peak in graphene \cite{PhysRevLett.97.187401}. 
By employing multiple excitation wavelengths, we access a pronounced resonance behavior when the laser energy is close to the excitonic transition of uncapped bismuthene \cite{Raul2022}. Under these resonant conditions, the intensity of the bismuthene-related mode increases significantly with respect to the SiC peaks and additional
phonon modes become visible, providing deeper insights into the possible scattering pathways in this material. 


The fabrication of this heterostructure starts with the growth of zero-layer graphene (ZLG) on SiC(0001), achieved through high-temperature silicon sublimation, which produces the characteristic $(6\sqrt{3} \times 6\sqrt{3})$R30$^\circ$ surface reconstruction \cite{Emtsev2009,Riedl2010}. Upon bismuth intercalation, the ZLG layer is effectively decoupled from the SiC substrate, lifting and converting it into a quasi-freestanding monolayer graphene \cite{Gehrig2025,Sohn2021,Tilgner2025,Stoehr2016}.

To first characterize the atomic structures of both the graphene overlayer and the underlying quantum material, we carried out \textit{in situ} scanning transmission electron microscopy (STEM) and scanning tunneling microscopy (STM) measurements. The experimental images and the resulting structural model are shown in Fig.~\ref{fig:fig1}a-c. STEM measurements (Fig.~\ref{fig:fig1}a) confirm the monolayer nature of bismuthene and reveal that the intercalated bismuth atoms occupy sites directly above the silicon atoms of the SiC substrate (T1 positions). By contrast, the graphene overlayer is only weakly resolved because of carbon's low atomic number, appearing solely as a subtle shoulder in the horizontally integrated line profile (white arrow) \cite{Schmitt2024NatComm,Gehrig2025}. Additional FFT-filtered constant-current STM measurements enable the individual layers to be selectively resolved at different bias voltages, revealing their respective in-plane atomic structures \cite{Schmitt2024NatComm,Gehrig2025}. At a sample bias of +50~mV, the graphene honeycomb lattice is imaged (Fig.~\ref{fig:fig1}b), whereas at $-400$~mV the tunneling signal is dominated by the bismuthene layer, revealing its honeycomb lattice (Fig.~\ref{fig:fig1}c). An extensive STM study of this system and the corresponding unprocessed images are available in Ref.~\cite{Gehrig2025}.

\begin{figure*}[t!]
\includegraphics[width=\linewidth, keepaspectratio]{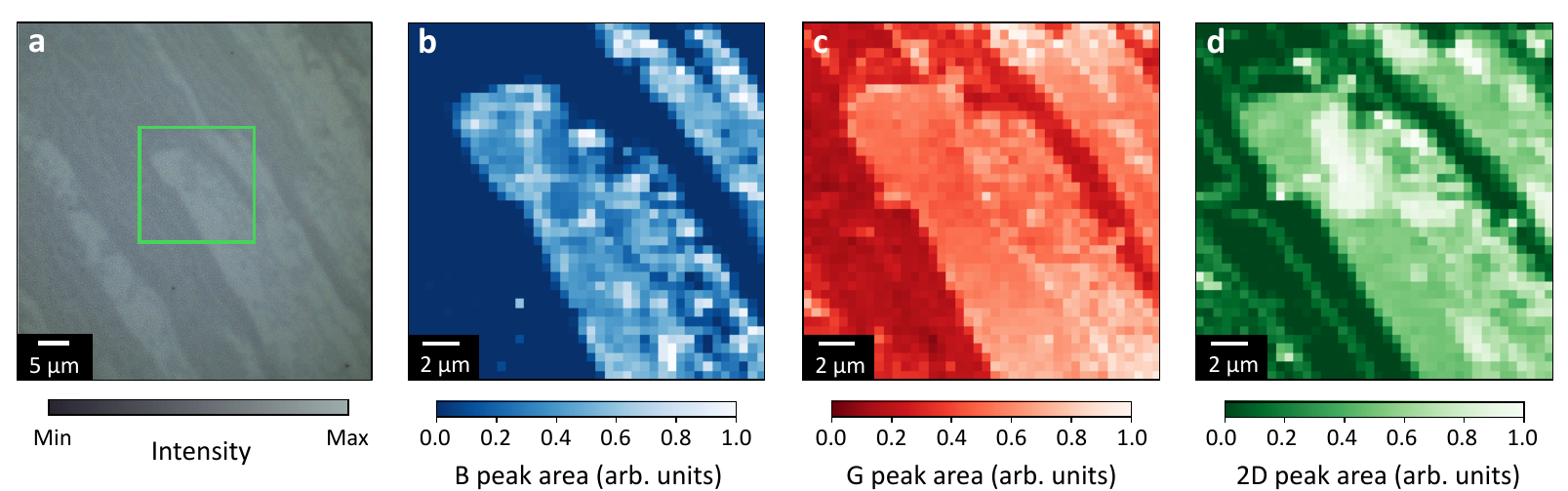}
\caption{\textbf{Spatial mapping of intercalated bismuthene} 
\textbf{a} Optical microscopy image of a sample, revealing two distinct phases: dark regions corresponding to non-intercalated areas and bright regions associated with intercalated domains. 
\textbf{b-d} Raman maps acquired within the area indicated by the green box in \textbf{a}, highlighting the spatial distribution of the respective spectral features. }
\label{fig:fig2}
\end{figure*}


Having established the structural properties of the system using UHV-based techniques, the graphene capping layer enables its investigation by \textit{ex situ} Raman spectroscopy. Raman measurements in Fig. \ref{fig:fig1}d were performed using an excitation energy of 2.33~eV and are shown for three representative spectral regions (blue, red, and green curve) of the graphene/bismuthene heterostructure. As a reference, Raman spectra were also acquired in regions where bismuthene and graphene Raman features were not expected (grey curves). These regions are mostly located at the edges of the same sample, where the large temperature gradient during growth suppresses graphene formation and, consequently, prevents bismuth intercalation. Throughout the following discussion, this reference is denoted as the bare SiC spectrum. We first examine the characteristic Raman features of graphene \cite{PhysRevLett.97.187401,MALARD200951}. The absence of the defect-induced D peak together with the exceptionally narrow 2D peak linewidth, with a full width at half maximum (FWHM) of approximately $26\,\mathrm{cm^{-1}}$, demonstrates the excellent crystalline quality of the graphene overlayer \cite{Neumann2015}. This linewidth is markedly reduced compared to graphene on SiC \cite{Lee2008} and other intercalation systems, underscoring the high structural integrity of the present sample \cite{MAMIYEV2025120002,SHTEPLIUK2019145}.

Beyond the characteristic graphene Raman modes, the spectra of the intercalated heterostructure exhibit an additional feature at 122~$\mathrm{cm}^{-1}$, which we attribute to the underlying bismuthene layer. This feature is absent in spectra from bare SiC and from graphene on SiC \cite{Wetherington_2021}, confirming its origin in the intercalant. To identify its microscopic origin, we performed density functional perturbation theory (DFPT) calculations to obtain the phonon dispersion of bismuthene on SiC, as shown in Fig.~\ref{fig:fig1}e. For simplicity, the calculations neglect the graphene capping layer, which, due to weak van der Waals bonding, has negligible influence on the phonon band structure \cite{Schmitt2024Raman,Ga2,Wetherington_2021}. The band colors indicate the character of each mode: green corresponds to pure bismuthene vibrations, black to SiC substrate modes, and red to mixed modes. Considering only the bismuthene contributions, the dispersion closely resembles that of graphene \cite{PhysRevB.71.205214,PhysRevB.67.035401}. 


Owing to the honeycomb lattice with two atoms per unit cell, both graphene and bismuthene possess three optical phonon branches. A doubly degenerate in-plane optical mode and a lower-energy out-of-plane optical mode are expected as first-order Raman modes ($q \approx 0$). In graphene, the doubly degenerate in-plane mode corresponds to the $E_{2g}$ phonon, giving rise to the prominent G peak at approximately 1580~cm$^{-1}$ \cite{PhysRevLett.97.187401,MALARD200951}. Likewise, the DFPT calculations predict a doubly degenerate zone-center optical phonon in bismuthene, giving rise to the pronounced Raman feature at approximately 122~cm$^{-1}$. By analogy with the graphene G peak, we refer to this mode as B peak. In addition, DFPT predicts a lower-energy zone-center optical mode of predominantly out-of-plane character at approximately 81~cm$^{-1}$, which is however Raman inactive, analogous to the silent low-frequency optical phonon in graphene \cite{Ferrari2013,NEMANICH1977}. This behavior is unique to honeycomb layers. In contrast, other intercalated 2D materials such as indium \cite{Schmitt2024Raman} or gallium \cite{Ga2}, which crystallize in a triangular lattice, do not exhibit distinct intrinsic optical modes. Their Raman response is instead dominated by weak interface modes arising from coupling between the intercalant layer and the substrate, in stark contrast to the pronounced intrinsic B peak of bismuthene.

The correspondence between the B and G peak is further corroborated by linearly polarized Raman measurements, in which the incoming polarization angle is kept fixed while the scattered one is rotated stepwise. As expected for a double-degenerate first-order zone-center mode, the B peak shows no polarization dependence, mirroring the behavior of the graphene G mode (Fig.~\ref{fig:fig1}f) \cite{yoon2008strong}. 

With the spectral fingerprints of the heterostructure established, we next examine their spatial distribution. The Raman signatures correlate directly with the optical contrast observed in microscopy images (Fig.~\ref{fig:fig2}a), where bright and dark regions correspond to intercalated and non-intercalated areas, respectively \cite{Schmitt2024Raman}. To quantitatively verify this assignment, spatially resolved Raman maps of the bismuthene B and graphene G and 2D modes were acquired (Fig.~\ref{fig:fig2}b--d). A strong spatial correlation between all three signals is observed. Regions exhibiting pronounced graphene G and 2D intensities simultaneously display the low-frequency bismuthene-related response, whereas these features are absent in the darker areas. The peak intensities were extracted by fitting the Raman modes with a Lorentzian at each spatial position. 
However, we note that certain regions escape this correlation, particularly at the boundaries of the bright areas, where graphene-related Raman features persist in the absence of the bismuthene B peak (see Supplementary Material). One possible explanation is local hydrogen intercalation induced during the high-temperature annealing step in hydrogen atmosphere, resulting in small patches of quasi-freestanding graphene adjacent to the bismuthene-intercalated regions. The coexistence of these two phases has previously been observed by ARPES \cite{Gehrig2025}. Additional contributions may arise from accelerated graphene growth at SiC step edges \cite{Emtsev2009,PhysRevB.80.121406}, where a graphene monolayer can form above the zero-layer graphene, thereby preventing bismuth intercalation \cite{Stoehr2016}. The characteristic bismuthene B peak is observed over the vast majority of the intercalated regions, demonstrating the robustness and uniformity of the bismuthene layer. The central frequency of the bismuthene B mode exhibits no significant spatial variation, suggesting a negligible level of strain gradients. This observation is consistent with the covalent bonding of bismuthene to the underlying SiC substrate. Moreover, the FWHM of the B peak is of about 5 cm$^{-1}$, constant throughout the map, suggesting a high crystalline order and a low number of structural defects. Beyond the individual Raman peak characteristics, the maps provide access to additional spectroscopic observables, such as the correlation between the spectral positions of the G and 2D bands ($\omega_G$–$\omega_{2D}$). Together, these indicate that the graphene is subject to a small residual strain and electron doping, in agreement with previous ARPES measurements \cite{Gehrig2025} (see SM for the extensive analysis of the Raman maps). This combined analysis establishes an all-optical quality-control approach for assessing the intercalation, extending from simple optical microscopy to Raman spectroscopy.

\begin{figure*}[t!]
\includegraphics[width=\linewidth, keepaspectratio]{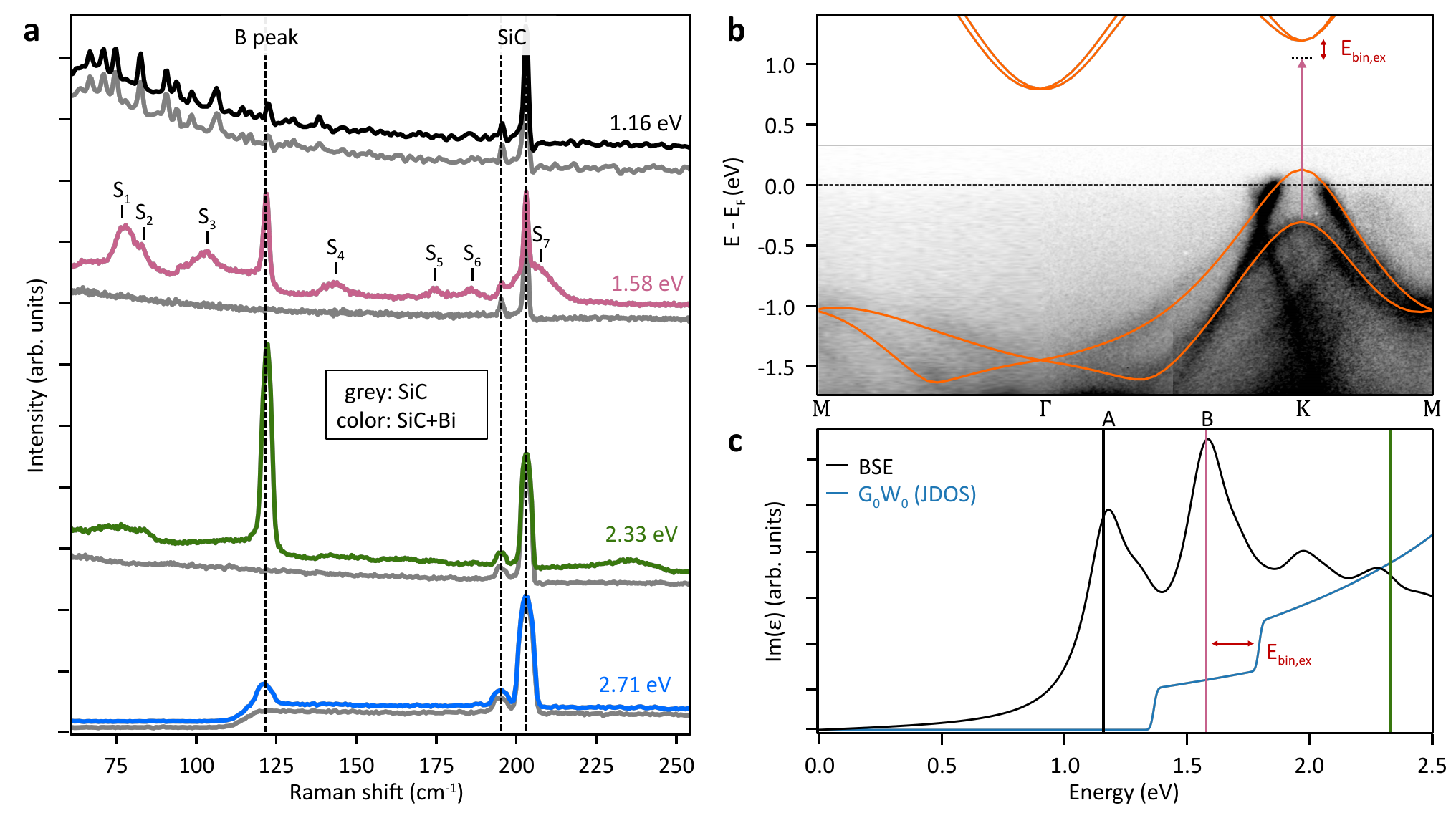}
\caption{\textbf{Resonantly-enhanced Raman response.} \textbf{a} Raman spectra measured at different excitation photon energies. \textbf{b} ARPES spectrum of intercalated bismuthene, in good agreement with DFT calculations (orange line), indicating slight $p$-type doping. The pink arrow marks the electronic transition associated with the B exciton and indicates its exciton binding energy $E_{\text{bin,ex}}$. \textbf{c} Absorption spectrum of pristine bismuthene calculated using the Bethe–Salpeter Equation (BSE), which accounts for the electron–hole Coulomb interaction and the resulting bound exciton states, is shown in black, while the GW-calculated joint density of states (jDOS) is shown in blue (taken from Ref.~\cite{Raul2022}). The step-like increases arise from direct optical transitions at the K point.}
\label{fig:fig3}
\end{figure*}

With the spatial distribution and crystalline quality of the heterostructure characterized, we now return to the Raman spectrum acquired from an intercalated region, where the characteristic B mode provides the vibrational fingerprint of bismuthene. Besides the B peak, the DFPT calculations shown in Fig.~\ref{fig:fig1}e also predict hybrid phonon modes involving both bismuthene and the SiC substrate. To verify the presence of these modes experimentally, the Raman spectrum acquired with an excitation energy of 2.33~eV is shown in Fig.~\ref{fig:fig3}a (green) for a larger spectral region. As discussed above, the dominant feature is the B peak at approximately 122~cm$^{-1}$. In addition, two pronounced peaks are observed near 200~cm$^{-1}$, which originate from the SiC substrate \cite{Schmitt2024Raman}. 
Besides this prominent features, two wide, faint peaks around 82~cm$^{-1}$ and 235~cm$^{-1}$ are also visible. The second one could be explained by a second order mode. To explain the other peak, we performed additional DFPT calculations using a simplified model system to include the graphene capping layer.
Because the complete intercalated structure corresponds to a $(6\sqrt{3}\!\times\!6\sqrt{3})$ supercell with respect to the primitive $(\sqrt{3}\!\times\!\sqrt{3})$ Bi/SiC unit cell, full phonon calculations are computationally impractical. Instead, we considered a $(2\times2)$ graphene layer on a $(\sqrt{3}\!\times\!\sqrt{3})$ Bi/SiC heterostructure, including van der Waals interactions and adjusting the lattice constant to preserve unstrained graphene. The calculations predict an interlayer breathing mode involving the bismuthene honeycomb lattice and the graphene overlayer at approximately 80~cm$^{-1}$, in good agreement with the experimentally observed weak feature at 82~cm$^{-1}$.

In Fig.~\ref{fig:fig1}e, several phonon modes predicted to originate from vibrations at the bismuthene--SiC interface remain unresolved, including the interlayer breathing mode expected near 100~cm$^{-1}$. To better characterize the Raman spectrum we perform frequency-dependent spectra aiming for excitation frequencies resonant with real electronic transitions \cite{cardona2005light,sotgiu2022raman}. We therefore first examine the electronic structure of the intercalated bismuthene by angle-resolved photoemission spectroscopy (ARPES), as shown in Fig.~\ref{fig:fig3}b. The ARPES band structure, overlaid with density functional theory (DFT) calculations, reveals a massive Dirac cone at the K point of bismuthene, accompanied by a pronounced Rashba splitting of the valence bands \cite{Gehrig2025}. In contrast to pristine bismuthene (without the graphene capping) \cite{Reis2017}, the graphene layer induces p-type doping, shifting the Fermi level below the upper valence-band maximum at K. Apart from this, the ARPES spectrum is in good agreement with the DFT calculations for pristine bismuthene. Additionally, the DFT calculations indicate an indirect band gap of approximately 0.8~eV between the K and $\Gamma$ points of the first Brillouin zone, as well as a direct band gap of about 1.2~eV at the K point. Previous studies on pristine bismuthene have shown that the near infrared optical absorption response of bismuthene is dominated by excitonic transitions \cite{Raul2022}, as reflected in the imaginary part of the dielectric function obtained from Bethe--Salpeter equation (BSE) calculations (Fig.~\ref{fig:fig3}c). A pronounced absorption feature, referred to as the A exciton, emerges near 1.2~eV, corresponding to the onset of the direct band gap reduced by the exciton binding energy. A second feature at approximately 1.63~eV, the B exciton, originates from the transition involving the lower valence band at the Dirac point. 

Guided by these excitonic resonances, we next investigate the Raman response under near-resonant excitation. We first employ a laser energy of 1.16~eV, which is nearly resonant with the A exciton. The resulting Raman spectrum is shown as the black curve in Fig.~\ref{fig:fig3}a, where the characteristic B mode is absent. This can be explained by the depopulation of the upper valence band in intercalated bismuthene, which suppresses the optical transition associated with the A exciton and hence the corresponding Raman resonance. Away from resonance, the Raman scattering intensity follows a fourth-power dependence on the excitation energy \cite{loudon1964raman}, making the intrinsic Raman modes effectively undetectable under infrared excitation. The Raman spectrum aquired at 1.16~eV shows additional features, which are also present in the bare substrate and could be explained by vibrational modes of ambient air \cite{Schmitt2024Raman}.
We therefore tune the excitation energy to 1.58~eV, close to the B exciton resonance. The corresponding spectrum is shown as the pink curve in Fig.~\ref{fig:fig3}a. A pronounced enhancement of the Raman response is observed, revealing several additional modes (S\textsubscript{1}–S\textsubscript{7}) alongside the B mode and the SiC doublet. We assign S\textsubscript{1} to the interlayer breathing mode involving the bismuthene honeycomb lattice and the graphene overlayer, which is already weakly visible under 2.33~eV excitation. S\textsubscript{3} is assigned to a breathing-like vibration of the bismuthene layer against the SiC substrate, consistent with the calculated phonon mode near 100~cm$^{-1}$ shown in Fig.~\ref{fig:fig1}e.
Beyond these first-order phonon modes, several additional spectral features (S\textsubscript{2}, S\textsubscript{4}-S\textsubscript{7}) emerge that cannot be attributed to single-phonon scattering processes. A comprehensive understanding of the origin of these additional spectral features would require temperature dependent measurements and the explicit evaluation of exciton--phonon coupling matrix elements \cite{nalabothula2025origin}, which lies beyond the scope of the present work. Nevertheless, these modes are most likely associated with higher-order, two-phonon scattering processes, analogous to observations reported for related systems such as indenene \cite{Schmitt2024Raman}, and become strongly enhanced under resonance condition. While a rigorous assignment of these second-order modes would require further theoretical investigation, comparison with the calculated two-phonon density of states suggests that double-resonant scattering pathways involving phonons near the \textbf{K} and \textbf{M} points of the first Brillouin zone dominate the Raman response \cite{guo2015double,sotgiu2026raman} (see Supplementary Materials for details).

Finally, we increase the excitation energy to 2.71~eV (blue curve in Fig.~\ref{fig:fig3}a). Although the B mode in the blue spectra remains discernible, its close proximity to the Rayleigh filter cut-off prevents a reliable analysis. Interestingly, no higher-order Raman modes are observed. This demonstrates that the Raman enhancement for 1.58~eV is governed by resonant optical absorption rather than by the increase in the GW-calculated joint density of states (jDOS) shown in Fig.~\ref{fig:fig3}c.



Since the excitation energy coincides with the excitonic transition of pristine bismuthene, we tentatively assign the observed Raman enhancement to an exciton-mediated scattering process. Even though the metallic graphene overlayer is expected to screen the Coulomb interaction, thereby modifying the exciton binding energy ($E_{\mathrm{bin,ex}}$) \cite{Raja2017}, 
the observation of a resonant Raman enhancement is nevertheless consistent with an excitonic contribution that is not completely suppressed by screening from the graphene overlayer. A similar partial reduction of the exciton binding energy by different numbers of graphene overlayer has been demonstrated for transition metal dichalcogenides, where dielectric screening weakens but does not eliminate the excitonic resonance \cite{Raja2017, tebbe2023tailoring,hill2017exciton}. A quantitative assessment of the influence of graphene on the excitonic properties of intercalated bismuthene, however, requires further investigation.

In conclusion, we have presented a comprehensive study of graphene-encapsulated bismuthene, combining STM and Raman micro-spectroscopy to reveal its structural and vibrational properties. A successful intercalation is clearly visualized in optical microscopy and confirmed via Raman spectroscopy. The identification of the first-order Raman B peak at $\sim$122~cm$^{-1}$, its polarization dependence, and its correspondence with theoretical phonon dispersions demonstrate a direct spectroscopic fingerprint of bismuthene. Across the mapped region, the B peak shows negligible spectral shifts and a linewidth of approximately 5 cm$^{-1}$, reflecting the high crystalline quality and a low defect density. The spatial correlation of graphene G and 2D modes with the bismuthene B peak provides a robust method to assess intercalation quality and sample uniformity. The observation of the B mode in \textit{ex situ} Raman measurements even after months provides further evidence for the environmental stability of intercalated bismuthene. With this approach, bismuthene is effectively \textit{capped and mapped}.

Furthermore, by tuning the excitation energy into resonance close to the excitonic transition in pristine bismuthene, we observe a pronounced enhancement of the Raman response, highlighting the role of resonant scattering processes. This resonant regime provides deeper insight into the coupling between electronic excitations and lattice dynamics.

These results establish a powerful and scalable approach for the characterization of graphene-capped quantum materials and open a door for exploring exciton-driven and topology-related phononic phenomena in bismuthene. Finally, owing to the symmetry of the bismuthene honeycomb lattice and its similarity in phonon band structure to graphene, one can anticipate the presence of topologically phonons, analogous to those in graphene \cite{Li2023}.

\bigskip
\textbf{Acknowledgements} 
We thank Armando Consiglio for the insightful discussions and for providing the BSE and GW-jDOS calculations. We are grateful for funding support from the Deutsche Forschungsgemeinschaft (DFG, German Research Foundation) under Germany's Excellence Strategy through the W\"urzburg-Dresden Cluster of Excellence on Complexity and Topology in Quantum Matter ctd.qmat (EXC 2147, Project ID 390858490) as well as through the Collaborative Research Center SFB 1170 ToCoTronics (Project ID 258499086). L.B. acknowledges financial support from the PNRR MUR Project (N. PE0000023-NQSTI).

\bigskip
\textbf{Data Availability} 
The data that support the ﬁndings of this study are available from the corresponding author upon reasonable request.


{\noindent
	\textbf{Author contributions} 
L.G. and C.Sc. have realized the sample growth and performed the scanning tunneling microscopy and photoelectron spectroscopy measurements. E.F., T.V. and S.S. conducted the Raman spectroscopy experiments. L.G., C.Sc., E.F., and S.S. performed the data analysis. S.E. have performed the DFPT  calculations. On the experimental side, contributions came from B.L., K.S., J.E., M.K., E.S., P.P., J.S., S.M., C.S., R.C. and L.B., while G.S. gave inputs to the theoretical aspects. L.B. supervised this joint project and wrote the manuscript together with all other authors.
}

{\noindent
	\textbf{Competing interests}
	The authors declare no competing interests.
}


%

\end{document}